# Performance Analysis of a 60 GHz Near Gigabit System for WPAN Applications

L. Rakotondrainibe, Y. Kokar, G. Zaharia, G. Grunfelder, G. El Zein
European University of Brittany (UEB), INSA, IETR - UMR CNRS 6164
INSA, 20 Avenue des buttes de Coesmes, CS 70839 -35708, Rennes cedex, France
lrakoton@insa-rennes.fr

*Abstract*—A 60 GHz wireless Gigabit Ethernet (G.E.) communication system capable of near gigabit data rate has been developed at IETR. The realized system covers 2 GHz available bandwidth. This paper describes the design and realization of the overall system including the baseband (BB), intermediate frequency (IF) and radiofrequency (RF) blocks. A differential binary shift keying (DBPSK) modulation and a differential demodulation are adopted at IF. In the BB processing block, an original byte/frame synchronization technique is designed to provide a small value of the preamble false alarm and missing probabilities. For the system performances, two different real scenarios are investigated: measurements carried out in a large gym and in hallways. Bit error rate (BER) measurements have been performed in different configurations: with/without RS (255, 239) coding, with frame synchronization using 32/64 bits preambles. As shown by simulation, the 64 bits preamble provides sufficient robustness and improves the system performance in term of BER. At a data rate of 875 Mbps, a BER of $10^{-8}$ was measured at 30 m using high gain antennas for line-of-sight (LOS) conditions.

*Keywords-Millimeter-wave system; WPAN; single carrier; BER; byte and frame synchronization*

## I. INTRODUCTION

60 GHz wireless systems, currently under standardization within the unlicensed 57-66 GHz band, are aiming several gigabits data rate for wireless personal area networks (WPANs) applications [1]-[3]. For any wireless system design, the selection of a modulation scheme is a main consideration and has a large impact on the system complexity. In fact, problems such as power amplifier (PA) non-linearity and oscillator phase noise are more important for these RF circuits resulting in performance degradation. These effects should be taken into account in the overall communication system. It was shown in [4] that single carrier (SC) transmission has a lower tolerance to phase noise and more resistant power PA non linearity than the multicarrier OFDM. Owing to these advantages, the authors in [5] proposed the single carrier (SC) transmission for multi-gigabit 60 GHz WPAN systems as defined in IEEE 802.15.3C standard. Up to now, in the literature, several studies have considered propagation measurements [6], [7], potential applications, circuit design issues and several modulations at 60 GHz [1]-[8]. However, few efforts have been dedicated to the realization of a 60 GHz wireless system and its performance in a realistic environment.

Due to the high path loss at 60 GHz and the transmission power restrictions, a simple solution is to use directional, high gain antennas. First of all, fading contributions are minimized by the spatial filtering effect of the antennas beamwidth, resulting in a higher coherence time. As shown in [6], when using directional antennas, the minimum observed coherence time was 32 ms (people walking at a speed of 1.7 m/s) which is much higher than the lower limit of 1 ms (omnidirectionnal antennas). Then, the channel is considered invariant during the coherence time and can be estimated once per few thousands of data symbols for Gbps transmission rate. Thus, the Doppler effect (particularly due to the moving person) depends on the antenna beamwidth but it is not considered critical in indoor environments. The use of directional antennas also yield the benefits of reducing the number of multipath components (the channel frequency selectivity) and therefore to simplify the signal processing. As stated in [2], [6], the root mean square (RMS) delay spread caused by multipath fading can be reduced to about 1 ns (the symbol duration for 1 Gbps with BPSK modulation). This means that the channel coherence bandwidth can be given as $B_{coh,\ 0.9} = 0.063/\tau_{rms} = 630$ MHz when using high gain antennas. In addition to a simple differential demodulation (which offers higher tolerance to the inter-symbol interference (ISI) than others SC modulations), the throughput less than 1 Gbps can be easily achieved without equalization. As the 60 GHz radio link operates only in a single room configuration, an additional Radio over Fibre (RoF) link is used to ensure the communications in all the rooms of a residential environment. For this reason, in this paper, we propose a hybrid optical/wireless system for the indoor gigabit WPANs. The first system application in a point-to-point configuration is the high-speed file transfer. Due to the cost of the transmission of the 60 GHz signals over RoF, it is reasonable to transmit signals over the fiber at IF.

This paper is organized as follows. Section II describes the transmitter (Tx) and the receiver (Rx). In this section, the baseband, the intermediate frequency and radiofrequency blocks are presented. In Section III, measurement results are presented; this section represents the core contribution of the paper. Section IV concludes the work.

## II. TRANSMITTER AND RECEIVER DESIGN

Fig. 1 and Fig. 2 show the block diagram of the Tx and Rx respectively. The multimedia data are transmitted from the source (video server) through the G.E. interface of the 60 GHz wireless transmitter.

This work is a part of Techim@ges research project supported by French "Media and Networks Cluster", Comidom and Palmyre II projects financed by the "Region Bretagne".

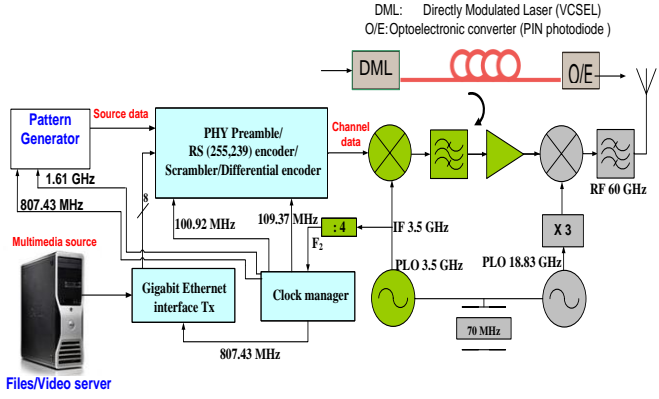

Figure 1. 60 GHz wireless Gigabit Ethernet transmitter

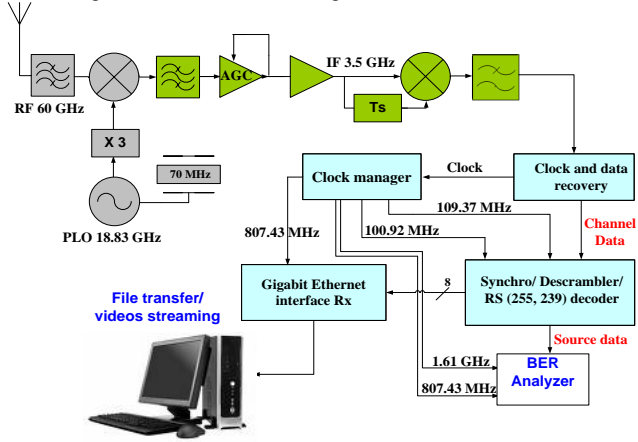

Figure 2. 60 GHz wireless Gigabit Ethernet receiver

The transmitted signal must contain timing information that allows the clock recovery and the byte/frame synchronization at the receiver (Rx) [5]. Thus, scrambling and preamble must be considered. A differential encoder allows removing the phase ambiguity at the Rx (by a differential demodulator).

*A. Transmitter design*

The Tx-G.E. interface is used to connect a home server to a wireless link with about 800 Mbps bit rate, as shown in Fig. 3. A header is inserted in the Ethernet frame to locate the starting point of each received Ethernet frame at the receiver.

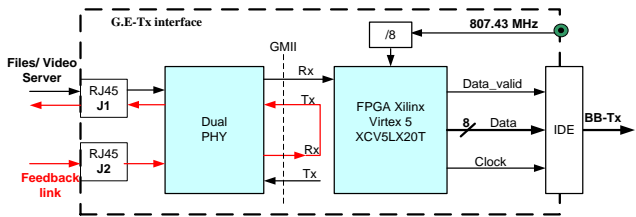

Figure 3. Transmitter Gigabit Ethernet interface

The gigabit media independent interface (GMII) is an interface between the media access control (MAC) device and the PHY layer. The GMII is an 8-bit parallel interface synchronized at a clock frequency of 125 MHz. However, this clock frequency is different from the source byte frequency $f_1$ = 807.43/8 =100.92 MHz generated by the clock manager in Fig. 1. Then, there is risk of packet loss since the source is always faster than the destination. In order to avoid the packet loss, a programmable logic circuit is used. The input byte stream is written into the dual port FIFO memory (FPGA) at a high frequency 125 MHz. The FIFO memory has been set up with two thresholds. When the upper threshold is attained, the dual PHY block (controlled by the FPGA) sends a 'signal stop' (to the multimedia source) in order to stop the byte transfer. A slow frequency $f_1$ reads out continuously the data stored in the FIFO. When the lower threshold is attained, the dual PHY block sends a 'signal start' to launch a new Ethernet frame. Therefore, whatever the activity on the Ethernet access, the throughput at the output of the G.E. interface is constant. Then, the byte stream from the G.E. interface is transferred in the BB-Tx, as shown in Fig. 4.

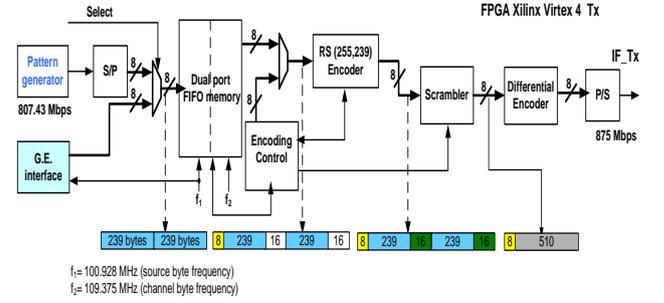

Figure 4. Transmitter baseband architecture (BB-Tx)

A known pseudorandom sequence of 63 bits is completed with one more bit to obtain an 8 bytes preamble. This preamble is sent at the beginning of each frame to achieve good frame synchronization at the receiver. Due to the byte operation of RS coding, two clock frequencies $f_1$ and $f_2$ are used:

$$f_1 = \frac{F_1}{8} = 100.929 \text{ MHz}, \quad f_2 = \frac{F_2}{8} = 109.375 \text{ MHz}. \quad (1)$$

where:

$$F_2 = \frac{3.5 \text{ GHz}}{4} = 875 \text{ MHz} \quad \text{and} \quad F_1 = \frac{2*239}{2*(239+16)+8} F_2.$$

As shown in Fig. 1, $F_2$ is obtained from the IF.

The frame format is realized as follows: the input source byte stream is written into the dual port FIFO memory at a slow frequency $f_1$. When the FIFO memory is half-full, the encoding control reads out data stored in the register at a higher frequency $f_2$. The encoding control generates an 8 bytes preamble at the beginning of each frame, which is bypassed by the RS encoder and the scrambler. The RS encoder reads one byte every clock period. After 239 clock periods, the encoding control interrupts the bytes transfer during 16 clock periods, so 16 check bytes are added by the encoder. In all, two successive data words of 239 bytes are coded before creating a new frame. After coding, the obtained data are scrambled using an 8 bytes scrambling sequence. The scrambling sequence is chosen in order to provide at the receiver the lowest false detection of the preamble from the scrambled data. Then, the obtained scrambled byte stream is differentially encoded before the modulation. The differential encoder performs the delayed modulo-2 addition of the input data bit ($b_k$) with the output bit ($d_k$):

$$d_{k+1} = d_k \oplus b_k \quad (2)$$

The obtained data are used to modulate an IF carrier generated by a 3.5 GHz phase locked oscillator (PLO) with a 70 MHz external reference. The IF signal is fed into a band-pass filter (BPF) with 2 GHz bandwidth and transmitted through a 300 meters fibre link. This IF signal is used to modulate directly the current of a laser diode operating at 850 nm. At the receiver, the optical signal is converted to an electrical signal by a PIN diode and amplified. The overall RoF link is designed to offer a gain of 0 dB.

The IF signal is sent to the RF block. This block is composed of a mixer, a frequency tripler, a PLO at 18.83 GHz and a band-pass filter (59-61 GHz). The local oscillator frequency is obtained with an 18.83 GHz PLO with the same 70 MHz reference and a frequency tripler. The phase noise of the 18.83 GHz PLO signal is about –110 dBc/Hz at 10 kHz off-carrier. The BPF prevents the spill-over into adjacent channels and removes out-of-band spurious signals caused by the modulator operation. The 0 dBm obtained signal is fed into the horn antenna with a gain of 22.4 dBi and a half power beamwidth (HPBW) of 10°V and 12°H.

### B. Receiver design

The receive antenna, identical to the transmit horn antenna, is connected to a band-pass filter (59-61 GHz). The RF filtered signal is down-converted to an IF signal centered at 3.5 GHz and fed into a band-pass filter with a bandwidth of 2 GHz. An automatic gain control (AGC) with 20 dB dynamic range is used to ensure a quasi-constant signal level at the demodulator input when the Tx-Rx distance or antenna moves. The AGC loop consists of a variable gain amplifier, a power detector and a circuitry using a baseband amplifier to deliver the AGC voltage. This voltage is proportional to the power of the received signal. A low noise amplifier (LNA) with a gain of 40 dB is used to achieve sufficient gain. A simple differential demodulation enables the coded signal to be demodulated and decoded. In fact, the demodulation, based on a mixer and a delay line (delay equal to the symbol duration $T_s$ = 1.14 ns), compares the signal phase of two consecutive symbols. A "1" is represented as a π-phase change and a "0" as no change, as in (2). Owing to the product of two consecutive symbols, the ratio between the main lobe and the side lobes of the channel impulse response increases. This means that the differential demodulation is more resistant to ISI effect compared to a coherent demodulation. Nevertheless, this differential demodulation is less performing in additive white Gaussian noise (AWGN) channel. Following the loop, a low-pass filter (LPF) with 1.8 GHz cut-off frequency removes the high-frequency components of the obtained signal. For a reliable clock acquisition realized by the clock and data recovery (CDR) circuit, long sequences of '0' or '1' must be avoided. Thus, the use of a scrambler (and descrambler) is necessary.

A block diagram of the baseband Rx is shown in Fig. 5. Owing to the RS (255, 239) decoder, the synchronized data from the CDR output are converted into a byte stream. Fig. 6 shows the architecture of byte/frame synchronization using a 64 bits preamble. The preamble detection is based on the cross-correlation of 64 successive received bits and the internal 64 bits preamble. Further, each $C_k$ (1 ≤ k ≤ 8) correlator of 64 bits must analyze a 1-bit shifted sequence. Therefore, the preamble detection is performed with 64+7 = 71 bits, due to the different possible shifts of a byte. In all, there are 8 correlators in each bank of correlators.

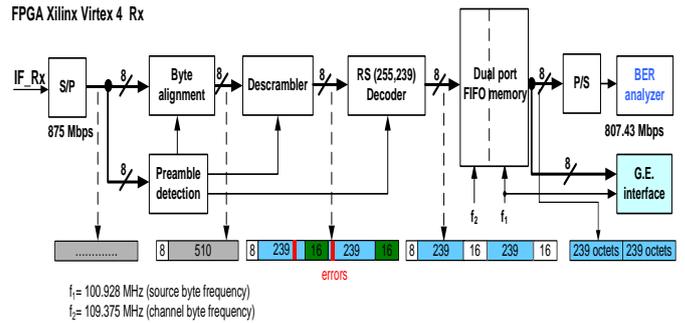

Figure 5. Receiver baseband architecture (BB-Rx)

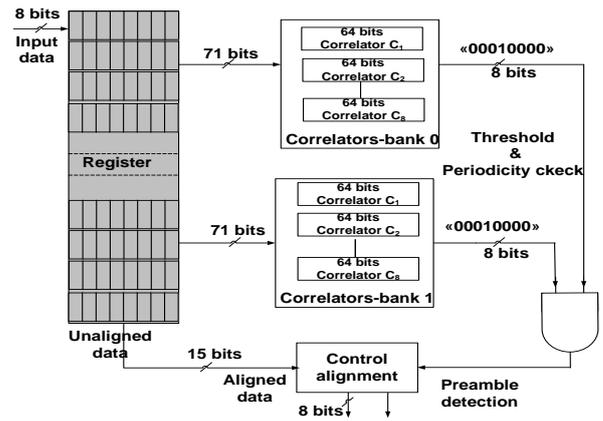

Figure 6. The preamble detection and byte synchronization

In addition, in order to improve the frame synchronization performance, two banks of correlators are used, taking into consideration the periodical repetition of the preamble: $P_1$ (8 bytes) + $D_1$ (510 bytes) + $P_2$ (8 bytes) + $D_2$ (510 bytes) + $P_3$ (8 bytes). This process diminishes the false alarm probability ($P_f$) while the missing detection probability ($P_m$) is approximately multiplied by 2, as shown later. The preamble detection is obtained if the same $C_k$ correlators in each bank of correlators indicate its presence. Therefore, the decision is made from 526 successive bytes ($P_1 + D_1 + P_2$) of received data stored by the receiving shift register. In fact, the value of each correlation is compared to a threshold ($\gamma$) to be determined. Setting the threshold at the maximum value ($\gamma$ = 64) is not practical, since a bit error in the preamble due to the channel impairments leads to a frame loss. A trade-off between $P_m$ and $P_f$ gives the threshold to be used. A false alarm is declared when the same $C_k$ correlators in each bank of correlators detect the presence of the preamble within the scrambled data ($D_1$ and $D_2$) [9].

The frame acquisition performance of the proposed 64 bits preamble was evaluated by simulations and compared to that of the 32 bits preamble [10]. The frame structure with 32 bits preamble uses only a data word of 256 bytes (255 bytes + a "dummy byte"). Fig. 7 and Fig. 8 show the missing probability versus channel error probability (p) and false alarm probability versus $\gamma$ for an AWGN channel, respectively.

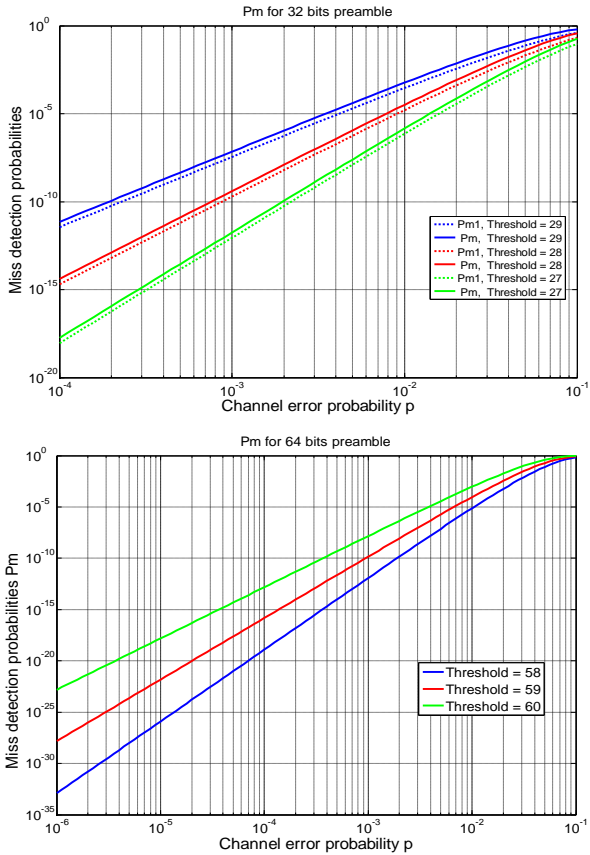

Figure 7. Miss detection probability with 32 bits and 64 bits preambles

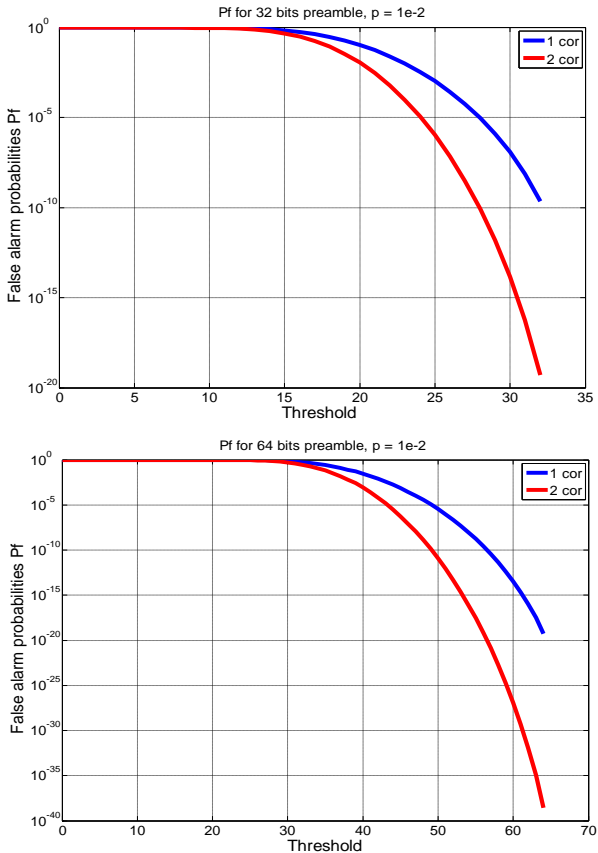

Figure 8. False alarm probability with 32 bits and 64 bits preambles

In these figures, $P_{m1}$ (or $P_{F1}$) and $P_m$ (or $P_{F2}$) indicate the missing (or false alarm) probabilities using one and two banks of correlators, respectively. The effect of p on the false alarm probability is insignificant since the random data bits "0" and "1" are assumed to be equiprobable.

With the 64 bits preamble, for $p = 10^{-3}$, the result indicate that $P_m = 10^{-10}$ and $P_{F2} = 10^{-24}$ for $\gamma = 59$. However, with the 32 bits preamble, we obtain $P_m = 10^{-7}$, $P_{F2} = 10^{-13}$ for $\gamma = 29$. This means that, for a data rate about 1 Gbps, the preamble can be lost several times per second because $P_m = 10^{-7}$ ($\gamma = 29$) with 32 bits preamble. We can notice that, for given values of p and $P_{F2}$, the 64 bits preamble shows a smaller missing probability compared to that obtained with the 32 bits preamble.

After the synchronization, the descrambler performs the modulo-2 addition between 8 successive received bytes and the descrambling sequence of 8 bytes. At the receiver, the baseband processing block regenerates the transmitted byte stream, which is then decoded by the RS decoder. The RS (255, 239) decoder can correct up to 8 erroneous bytes and operates at a fast clock frequency $f_2 = 109.37$ MHz. The byte stream is written discontinuously into the dual port FIFO memory at a fast clock frequency $f_2$. A slow clock frequency $f_1 = 100.92$ MHz reads out continuously the byte stream stored by the register, since all redundant information is removed. Afterwards, the byte stream is transferred to the receiver Gigabit Ethernet interface, as shown in Fig. 9. The feedback signal can be transmitted via a wired Ethernet connection or a Wi-Fi radio link due to its low throughput.

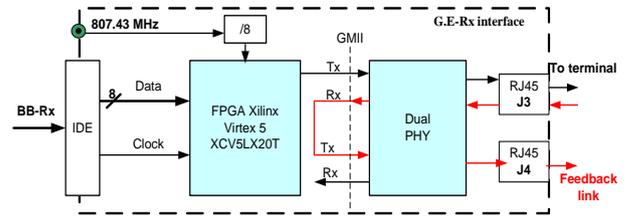

Figure 9. Receiver Gigabit Ethernet interface.

### III. MEASUREMENT RESULTS

Back-to-back test of the realized system (without RF and AGC loop) was firstly carried out. The goal is to evaluate the BER of the realized system versus the signal to noise ratio (SNR) at the demodulator input. An external AWGN is added to the IF modulated signal (before the IF-Rx band pass filter). The external AWGN is a thermal noise generated and amplified by successive amplifiers. This noise feeds a band pass filter and a variable attenuator so that the SNR is varied by changing the noise power. The BER versus SNR is shown in Fig. 10.

Compared to the theoretical performance, for $BER = 10^{-5}$, the SNR degradation of the realized system is about 3.5 and 3 dB for uncoded and coded data, respectively. This degradation is mainly due to the 2 GHz available bandwidth. This bandwidth is too wide for a throughput of 875 Mbps. In order to avoid the increased power noise in the band, the filter bandwidth must be reduced to 1.1 GHz (a roll-off factor 0.25) [5]. Using the free space model, Fig. 11 shows the estimated IF received power versus the Tx-Rx distance. This result takes into account the transmitted power, the antenna gains, the path loss and the implementation losses of RF blocks. Two types of antennas were used: horn antenna and patch antenna. The patch antenna has a gain of 8 dBi and a HPBW of 30°. The IF receiver noise level is:

$$N_L = -174 \text{ (dBm/Hz)} + NF + 10\log(B) = -71.98 \text{ dBm} \quad (3)$$

where NF = 9 dB is the total noise figure and $B = 2*10^9$ Hz is the receiver bandwidth. As shown in Fig. 10, the minimum SNR needed for BER = $10^{-4}$ is about 10.5 dB. Thus, the receiver sensitivity is about S = - 61.5 dBm. Moreover, the demodulator input power must be greater than 0 dBm (due to the conversion loss of the power splitter and the minimum power level required at the mixer input). These values give an indication of the maximum distance accepted by the system.

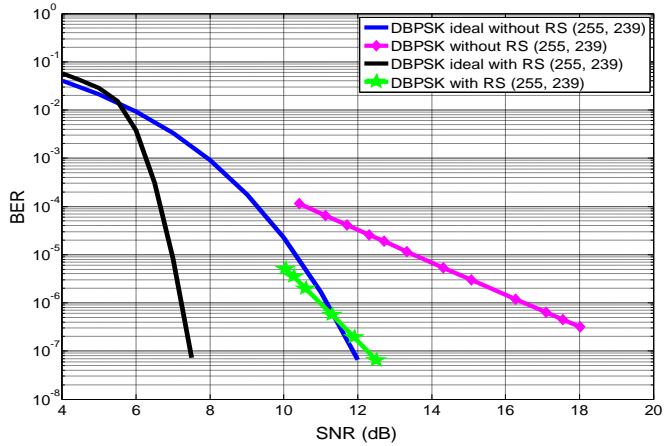

Figure 10. BER versus SNR including AWGN channel

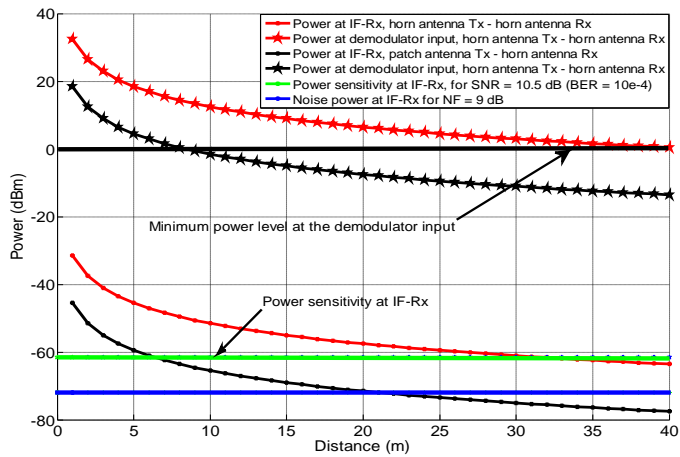

Figure 11. The IF received power versus Tx-Rx distance

BER measurements were performed over distances ranging from 1 to 40 meters in a large gym. At each distance, the BER was recorded during 10 minutes. The Tx and Rx horn antennas (placed in the middle of the empty gym) were situated at a height of 1.35 m above the floor. These measurements were conducted under LOS conditions with a fixed Rx and the Tx placed on a trolley pushed in a horizontal plane to various points about the environment. During the measurements, the Tx and Rx were kept stationary, without movement of persons. A PN sequence of 127 bits provided by a pattern generator was used as data source. As shown in Fig. 12, measured BER without BB blocks were obtained at the CDR output for two possible configurations: AGC minimum and maximum gains. In this case, the AGC loop is disconnected but an external DC voltage controls the AGC amplifier. As a result, for BER around $10^{-4}$, when the AGC amplifier is set at a gain of 8 dB, the upper limit of the Tx-Rx distance is about 7 m whereas this maximum distance can be increased at 35 m, using an AGC amplifier with 28 dB gain. The BER result with RS (255, 239) coding is also shown in Fig. 12 (frame structure using 32 bits preamble for a threshold $\gamma = 29$). It is shown that for a BER at $10^{-6}$, the distance is equal to 27 m without channel coding and 36 m with RS coding. This result proves the RS coding efficiency. Compared to the predicted distance obtained in Fig. 11, using high gain antennas, measured propagation loss characteristics showed very good agreement. The effects of multipath propagation on the BER performance are greatly reduced by the spatial filtering of the directive horn antennas.

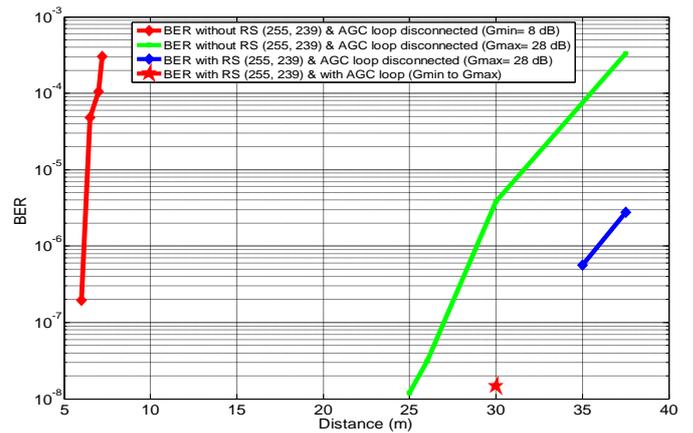

Figure 12. BER versus the Tx-Rx distance (32 bits preamble)

The main problem of using directional antennas is the human obstruction. The signal reaching the receiver is randomly affected by people moving in the area and can lead to frequent outages of the radio link. For properly aligned antennas, it is confirmed that the communication is entirely interrupted when the direct path is blocked by a human body (synchronization loss during the obstruction duration). The measurement results show a further attenuation of around 20 dB on the received power (as indicated in the power detector) when a person moves in front of the receiver. High gain antennas are needed for the 60 GHz radio propagation but to overcome this major problem, it is possible to exploit the angular diversity obtained by switching antennas or by beamforming [5]. To improve the system reliability, a Tx mounted on the ceiling, preferably placed in the middle of the room can mitigate the radio beam blockage caused by people or furniture [7], [8]. In real applications, the Tx antenna should have a large beamwidth to cover all the devices operating at 60 GHz in a room and the Rx antenna placed within the room should be directive so that the LOS components are amplified and the reflected components are attenuated by the antenna pattern.

In order to examine the effects of the antenna directivity and the multipath fading, BER measurements were also conducted within hallway over distances ranging from 1 to 40 meters, as shown in Fig. 13. The door of a 4 cm thickness (agglomerated wood), was opened during the BER measurements. The hallway has concrete walls and wooden doors on both sides. The Tx-Rx antennas (placed in the middle of the hallway) were positioned at a height of 1.35 m. The idea was to analyze the results of BER measurements with and without RS coding in a wider hallway separated by a door.

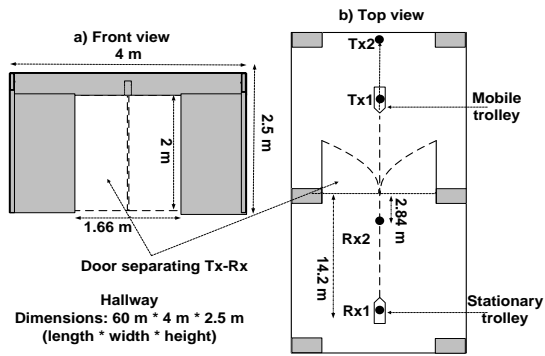

Figure 13. The hallway: a) front view; b) top view

Due to the guided nature of the radio propagation along the hallway, the major part of the transmitted power is focused in the direction of the receiver. However, the Rx horn antenna needs to be properly well aligned in the direction of the Tx, the misalignment of antennas (of a few degrees) results in a significant degradation of the LOS component and increases the multipath components caused by the sides of walls and the door borders. In a large gym, we found that the beam misalignment of directive antennas (of a few degrees) is less critical in term of BER degradation (distance less than 20 m). In the worst case, in hallway or in a large gym, the misalignment errors can lead to occasional synchronization losses, giving a BER higher than $10^{-4}$.

In the hallway, the door and walls can cause reflections and diffractions of the transmitted signal, in particular when the Rx position is far away from the opening door (as Rx1 position shown in Fig. 13). We found that for the same 32 m Tx-Rx distance, the received signal power was similar for both positions Tx1-Rx1 and Tx2-Rx2. However, the BER without coding is equal to $2.8*10^{-2}$ (synchronization loss) and $2.8*10^{-5}$ for Tx1-Rx1 and Tx2-Rx2 positions, respectively. In the case of Tx1-Rx1 position, diffractions and reflections from the borders of the opening door can be the dominant contributors to the significant BER degradation.

BER measurements versus Tx-Rx distance using 64 bits preamble were also carried out (Rx2 position). We found that for a Tx-Rx distance less than 32 m, the number of errors with RS coding are equal to 0 (the BER was recorded during 5 minutes), using 64 bits preamble ($\gamma = 58$). Compared to the result obtained with the 32 bits preamble, as shown in Fig. 12, our investigation revealed that the proposed 64 bits preamble improves the system robustness in a realistic environment.

We also evaluate the BER performance when the door is closed. We observed that the attenuation increases of about 15 dB. A similar value was obtained in [7]. In this situation, the propagation channel is also unavailable during the "shadowing events" and lead to permanent synchronization loss. Therefore, radio electric openings (windows ...) are necessary.

IV CONCLUSIONS

This paper shows a full experimental implementation of a 60 GHz unidirectional wireless system. The proposed system provides a good trade-off between performance and complexity. An original method used for the byte/frame synchronization is also described. The numerical results show that the proposed 64 bits preamble allows obtaining better BER results comparing to the previously proposed 32 bits preamble [10]. This new frame format allows obtaining a high preamble detection probability and a very small false alarm probability. As a result, a Tx-Rx distance greater than 30 meters was reached with low BER using high gain antennas. Our investigation revealed that the high gain antenna directivity stresses the importance of the antennas pointing precision. In order to support a Gbps reliable transmission within a large room and severe multipath dispersion, a convenient solution is to use high gain antennas. However, the use of directional antennas for 60 GHz WPAN applications is very sensitive to objects blocking the LOS path.

Due to the hardware constraints, the first data rate was chosen at 875 Mbps. Using a new CDR circuit limited at 2.7 Gbps, a data rate of 1.75 Gbps can be achieved with the same DBPSK architecture or with DQPSK architecture. For suitable quality requirements in Gbps throughput, an adaptive equalizer should be added to counteract the ISI influence.